\documentclass[a4paper]{article}

\usepackage{a4wide}
\usepackage{graphicx}
\usepackage{hyperref}
\usepackage{amsmath}
\usepackage{parskip}

\begin{document}
\title{Tautochrone and Brachistochrone Shape Solutions for Rocking Rigid Bodies }
\author{Patrick Glaschke}
\maketitle

\begin{abstract}
Rocking rigid bodies appear in several shapes in everyday life: As furniture like
rocking chairs and rocking cradles or as toys like rocking horses or tilting dolls.
The familiar rocking motion of these objects, a non-linear combination of
a rigid rotation and a translation of the center of mass, gives rise to a number
of interesting dynamical properties. However, their study has received little
attention in the literature.

This work presents a comprehensive introduction to the dynamics of rocking rigid bodies,
including a concise derivation of the equations of motion as well as a general inversion
procedure to construct rocking rigid body shapes with specified dynamical properties.
Moreover, two novel rigid body shapes are derived --- the tautochrone shape
and the brachistochrone shape --- which represent an intriguing generalization of the
well-know tautochrone and brachistochrone curves. In particular, tautochrone
shapes offer an alternative construction of a tautochrone pendulum, in addition
to Huygens' cycloid pendulum solution.

\vspace*{3mm}

\end{abstract}

\section{Introduction}

Rocking rigid bodies exhibit a rich dynamic behavior: They rock and roll, slide or tumble over
and might deadlock at concave boundary sections. The governing equations of motion are strongly
non-linear, such that their study requires numerical studies or semi-analytical techniques. As
such, they have been widely studied in the field of earth quake safety analysis \cite{aslam2011rocking,iyengar1991rocking,ishiyama1982motions,nozaki2009study,schau2013rocking,winkler1995response}.
In addition, detailed studies of static equilibria of balanced rigid bodies established close
relations to geometrical and topological theorems \cite{domokos2012mechanics}. The quest for minimizing
the number of equilibrium points resulted in the discovery of the G\"omb\"oc, a remarkable three-dimensional
shape featuring only one stable and one unstable equilibrium \cite{varkonyi2006static}.
Despite these diverse research approaches there are two classical mechanics problems which have not been
applied to rocking rigid bodies yet: the
tautochrone\footnote{From Ancient Greek, meaning ``same time''. ``isochrone'' is in use as well.}  problem and the
brachistochrone\footnote{From Ancient Greek, meaning ``shortest time''.}  problem. The tautochrone problem asks for the path along which a frictionless gliding bead returns in constant time to a fixed reference position, independent of the starting point on that path.
Christiaan Huygens already found  in 1660 the solution for a bead moving in the homogeneous gravity field of the
earth: the inverted cycloid \cite{emmerson2005things}. The brachistochrone problem asks for the path along which a frictionless bead
slides from a given
starting point to a given end point in minimum time. Though both problems seem to be unrelated, Johann Bernoulli discovered later
in 1697 that the brachistochrone curve in an homogeneous gravity field is given by the same inverted
cycloid\footnote{The coincidence of tautochrone and brachistochrone paths does not apply to all potentials, see \cite{denman1985remarks}
for a detailed account.}. As the calculation of the brachistochrone path requires the minimization of an integral, it also initiated
a whole new field, the calculus of variations, and has seen many generalizations since. Quickest paths of descent have been derived for
other potentials, geometries, curved space-time and various frictional forces
\cite{gemmer2006generalizations,jeremic2011brachistochrone,legeza2010brachistochrone,mertens2008brachistochrones,parnovsky1998some,prieto2005rocking}. The tautochrone
problem has seen similar generalizations as well \cite{flores1999tautochrone,kamath1992relativistic,munoz2010tautochrone}.

\begin{samepage}
As none of these generalizations covers rocking rigid bodies, let me begin with a close adaptation
of both problems to planar (i.e.\ slab-like) rocking rigid bodies:
\begin{itemize}
\item A rocking rigid body is said to have a {\em tautochrone shape} if it returns to its equilibrium position in constant time.
\item A rocking rigid body is said to have a {\em brachistochrone shape} if it returns from a given initial orientation to its
equilibrium position in minimal time.
\end{itemize}
\end{samepage}
Having stated the problems to be solved, the remainder of the paper addresses the derivation of both novel shape solutions.
The account starts with an introduction to rocking rigid body dynamics using complex calculus, followed by a generic inversion procedure
to derive rigid body shapes from any desired rocking characteristics. Then both shapes solutions are derived, including a brief analysis
of the balance of frictional forces at the contact point to explore the actual construction of these shapes.
 
\section{Rocking Rigid Bodies}

\begin{figure}
\centering
\includegraphics*[scale=.5, angle=270]{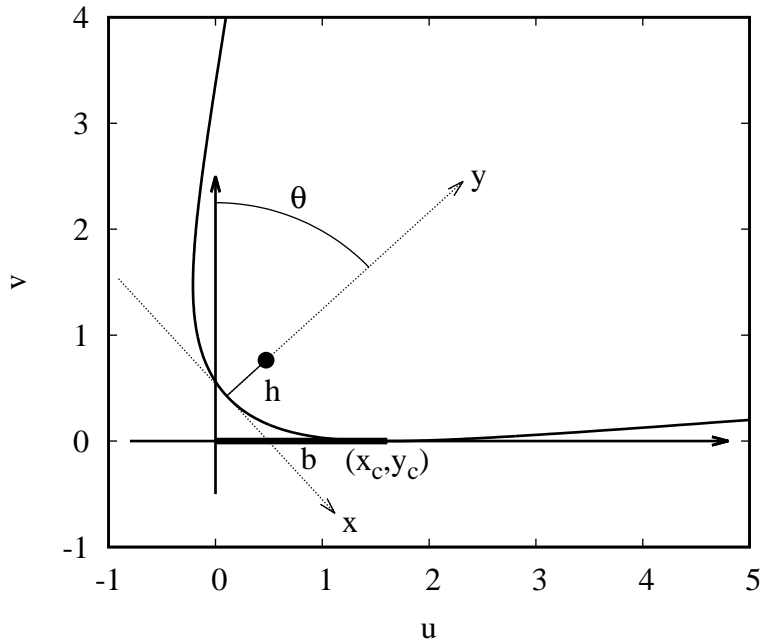}
\caption{\label{StehA}Rocking rigid body and used coordinate system. Unrolled curve length $b$ and center of mass
are highlighted as well. The shape function is the same as in Fig.~\ref{Demo}.}
\end{figure}

The study of a rocking rigid body requires an accurate description of the rigid body and its dynamics.
It is customary to use two coordinate systems: A body-fixed coordinate system ($x$,$y$ in the following) and a space-fixed
coordinate system ($u$,$v$ in the following). The body rolls along a rigid support plane defined at $v=0$. Without loss of
generality, the body-fixed coordinate system is defined such that it coincides with the space-fixed coordinate system when
the rigid body assumes some (preferably stable) equilibrium position. In particular, the origin is chosen such that it coincides with the
corresponding equilibrium contact point.

The mass distribution of a rigid body is sufficiently described by its moment of inertia $\Theta$ and the position of its center of mass,
located at $(0,h)$ owing to the definition of the body-fixed coordinate system. For the time being, I will assume a rocking motion without
slip. In virtue of that assumption, any rotation by an angle $\theta$ (measured in clockwise direction) induces a translation
$b$ in $u$-direction. $b$ is defined as the unrolled curve length of the body contour measured from the origin
to the contact point $(x_c, y_c)$. $\theta$ and $b$ are related to the local radius of curvature $r$ by
\begin{eqnarray}
r &=& \frac{db}{d\theta}\,. \label{def_r}
\end{eqnarray}
Negative radii correspond to a ``hanging'' rigid body, that is the rigid body rolls along the underside of the support plane.
The contour traced out by all contact points is the shape function of the rigid
body, or simply its shape. Fig.~\ref{StehA} summarizes all introduced variables
and coordinate systems. The transformation from body-fixed coordinates $(x',y')$ to space-fixed coordinates $(u',v')$ is given by
\begin{eqnarray}
\left( \begin{array}{c} u' \\ v' \end{array} \right)  &=&  \left( \begin{array}{c} b \\ 0 \end{array} \right)
+ \left( \begin{array}{rr} \cos(\theta) & \sin(\theta) \\ -\sin(\theta)  & \cos(\theta) \end{array} \right)
\left( \begin{array}{r} x' - x_c \\ y' - y_c \end{array} \right) \label{xytouv} \\
\left( \begin{array}{c} \cos(\theta) \\ \sin(\theta) \end{array} \right) &=& \frac{d}{db}
\left( \begin{array}{c}  x_c \\ y_c \end{array} \right)\,. \label{def_theta}
\end{eqnarray}
Though only convex rigid bodies are actually capable of rolling along a planar surface, Eq.~\ref{xytouv} extends the notion of ``rolling''
to any shape function where $\theta$ is a continuous function on the curve. Interestingly, this condition does not imply that a shape
function needs to be differentiable everywhere. For instance, a figure-3 shaped curve is not differentiable at the join of the two arcs, yet a
rolling procedure according to Eq.~\ref{xytouv} is still well defined. The rolling motion of a rigid body is a special case of a much more
general geometrically operation know as {\em roulette} \cite{besant1890notes}. A roulette is defined as the curve traced out by a reference
point attached to a curve which rolls along a second, fixed reference curve.

Having thus established a generalized definition of rolling, it
is worth to have a closer look at the parametrization of the rocking motion itself. The orientation of the rigid body might either be parametrized
by the rocking angle $\theta$, the unrolled curve length $b$ or the height of the center of mass $v$. However, none of these parameters is
guaranteed to uniquely parameterize the rocking motion, as they are generally related by multi-valued functions. Therefore a careful restriction of
the parameter domains is required to obtain meaningful results for the problem at hand.

Using Eq.~\ref{xytouv} to calculate the center of mass path $(u,v)$ in
space-fixed coordinates gives
\begin{eqnarray}
\left( \begin{array}{c} u \\ v \end{array} \right)  &=&  \left( \begin{array}{c} b \\ 0 \end{array} \right)
+ \left( \begin{array}{rr} \cos(\theta) & \sin(\theta) \\ -\sin(\theta)  & \cos(\theta) \end{array} \right)
\left( \begin{array}{r}  - x \\ h - y \end{array} \right)\,.
\end{eqnarray}
As $(x,y)$ refers exclusively to the contact point in the following, the subscript $c$ is dropped from now for brevity.
The transformation described by Eqs.~\ref{xytouv}--\ref{def_theta} suggests the introduction of complex
coordinates\footnote{Another interesting application is to express the rocking angle as a complex number to solve the equations of motion, see \cite{prieto2005rocking} for details.} $z = x+iy$, $w = u+iv$ to simplify the center of mass equation to
\begin{eqnarray}
w_b &=& \frac{dz^*}{db}(ih-z) \label{Eqdz_wb} \\
w_b &:=& w-b \\
\frac{dz}{db} &=& \exp(i \theta)\,. \label{Eqdz_db}
\end{eqnarray}
$w_b$ is the center of mass position measured relative to the contact point. Using the notion introduced above,
the center of mass path $w$ is the roulette generated by the support plane, the shape function and the center of mass.
For a better overview, Table~\ref{UsedCoord} contains a summary of the used variables.

Useful is the relation
\begin{eqnarray}
\frac{dw}{d\theta} &=& -i \frac{dz^*}{db} (ih-z) \\
&=&  -i w_b  \label{w_phiDefEq}
\end{eqnarray}
which demonstrates that the contact point is the instantaneous center of rotation. By
considering the imaginary part only it follows that
\begin{eqnarray}
\frac{dv}{d\theta} &=& - u_b \\                         \label{dv_dtheta}
&=&\frac{1}{2} \frac{d |w_b|^2}{d b} \label{dv_dwb_db}
\end{eqnarray}
which can be rearranged to provide an alternative expression for the local radius of curvature:
\begin{eqnarray}
r &=&  \frac{1}{2}  \frac{d|w_b|^2}{d v} \label{r_dwb_db}\,.
\end{eqnarray}
Finally, by using the real part of  Eq.~\ref{w_phiDefEq} we obtain a differential equation for the path of the center of mass
in space-fixed coordinates:
\begin{eqnarray}
\frac{d u}{d v}     &=& v \frac{d\theta}{d v}\,. \label{du_dv}
\end{eqnarray}
The general analysis itself presented in this section allows for many interesting applications. As the primary focus
of this work is a study of tautochrone and brachistochrone shapes, an illustrative example using conic sections can
be found in Appendix~\ref{AppendixConical}.

\begin{table}
\centering
\begin{tabular}{|l|c|c|c|} \hline
& body-fixed  & space-fixed & comoving \\ \hline
Center of mass &  $ih$       & $w$         & $w_b$    \\ \hline
Contact point &  $z$        & $b$         & 0 \\ \hline
\end{tabular}
\caption{\label{UsedCoord}Used coordinate systems.}
\end{table}

\section{Static Equilibrium}

A rocking solid in static equilibrium does not experience a net torque. This requires the center of mass to be vertically aligned
with the contact point and $u_b=0$ holds. For smooth shapes this is equivalent to
\begin{eqnarray}
\frac{dv}{d\theta} &=& 0\,. \label{eqilibrium_eq}
\end{eqnarray}
Furthermore, following from Eq.~\ref{dv_dwb_db}, equilibrium points are always stationary points of the distance from the center of mass to the
contact point. Planar rigid bodies of finite extend have at least one stable and one unstable equilibrium point, which is a direct consequence
of the maximum-minimum theorem. Bodies with only one stable equilibrium are called {\em monostatic}, while the special case of one
stable and one unstable equilibrium is denoted as {\em mono-monostatic}. A monostatic body is quite easily constructed, unless
its mass distribution is required to be homogeneous. It can be shown that two-dimensional homogeneous monostatic bodies do not exist, which is
equivalent to the famous Four-Vertex theorem \cite{varkonyi2006static}.
 
Differentiating Eq.~\ref{eqilibrium_eq} with respect to $\theta$ provides further insight into the type of equilibrium.
The derivative is
\begin{eqnarray}
\frac{d^2v}{d\theta^2}  &=& r - v
\end{eqnarray}
which proves the following physical classification of equilibria:
\begin{itemize}
\item An equilibrium is stable if the center of mass is located below the local center of curvature
\item An equilibrium is neutral if the center of mass coincides with the local center of curvature
\item An equilibrium is unstable if the center of mass is located above  the local center of curvature
\end{itemize}
These conditions apply equally well to negative radii of curvature, though this requires a generalized rocking motion
in the sense of Eq.~\ref{xytouv}.

\section{Equations of Motion}

The motion of a rocking body combines a translation of the center of mass with
a rotational motion in the gravity field of the earth. Thus the Lagrangian (normalized by the mass $m$ of the rigid body) reads
\begin{eqnarray}
\mathcal{L} &=& \frac{1}{2}\left( {\dot u}^2 + {\dot v}^2\right) + \frac{1}{2} \Theta \, {\dot \theta}^2 - g  v
\end{eqnarray}
where dots indicate the derivative with respect to time.
$\Theta$ is the normalized moment of inertia (equal to the squared radius of gyration) and $g$ is the gravitational acceleration of
the earth. In virtue of the no-slip assumption, both $u$ and $v$ can be considered as functions of $\theta$ only which further
simplifies the Lagrangian to:
\begin{eqnarray}
\mathcal{L} &=& \frac{1}{2}\left( |w_b|^2 + \Theta \right) {\dot \theta}^2 - g  v\,.
\end{eqnarray}
The Euler-Lagrange equation of motion is:
\begin{eqnarray}
(|w_b|^2+\Theta) \ddot \theta                           &=&  {\dot \theta}^2 u_b r - g \frac{dv}{d\theta} \,. \label{EqofMot}
\end{eqnarray}
Before examining the exact solution of Eq.~\ref{EqofMot}, it is useful to consider the small angle approximation
\begin{eqnarray}
(h^2+\Theta) \ddot \theta    &=&   - g (r_0 -h ) \theta\,.
\end{eqnarray}
This is the differential equation of an harmonic oscillator with frequency
\begin{eqnarray}
\omega_0^2 &=& \frac{g(r_0-h)}{h^2+\Theta}. \label{eq_w0}
\end{eqnarray}
For later use, it is convenient to introduce the half-length $L$ of a pendulum  with frequency $\omega_0$:
\begin{eqnarray}
L &:=& \frac{h^2+\Theta}{2(r_0-h)}\,.
\end{eqnarray}
To solve the full equation of motion, I introduce a new auxiliary parameter $s$
\begin{eqnarray}
s(\theta)&:=&\int_0^{\theta}\sqrt{ |w_b|^2+\Theta} \, d\theta \label{sDefEq}
\end{eqnarray}
which simplifies the Lagrangian and the equation of motion to:
\begin{eqnarray}
\mathcal{L} &=& \frac{1}{2} {\dot s}^2 - g v(s) \\
\ddot s &=& - g \frac{dv}{ds} \label{s_diff_eq} \,.
\end{eqnarray}
Hence the dynamics of a rocking rigid body is equivalent to a point mass moving in an effective potential $U(s) = gv(s)$.
Eq.~\ref{s_diff_eq} can be solved by invoking the conservation of energy:
\begin{eqnarray}
t - t_0 &=& \int_{s_0}^{s(t)} \frac{ds'}{\sqrt{2g(v_{A}-v(s'))} } \,.
\end{eqnarray}
$v_{A}$ sets the maximal amplitude of the rocking motion.

\section{Constraint Force \label{SecConstForce}}

The weight of the rocking rigid body as well as the dynamic forces due to the rocking motion
need to be balanced by the support plane. Therefore the constraint acceleration acting on the contact point
is
\begin{eqnarray}
a_c &=& \ddot w  + ig \,.
\end{eqnarray}
Applying Eq.~\ref{w_phiDefEq} twice provides the velocity and acceleration of the center of mass:
\begin{eqnarray}
\dot w &=& -i \dot  \theta w_b \\
\ddot w &=& -i \ddot \theta w_b + (i r - w_b) {\dot \theta}^2  \,.
\end{eqnarray}
Inserting Eq.~\ref{EqofMot} gives the desired acceleration
\begin{eqnarray}
a_c    &=& \left( {\dot \theta}^2  r + g \right) \frac{v w_b + i\Theta}{|w_b|^2+\Theta}  - w_b {\dot \theta}^2 \,.
\end{eqnarray}

\section{Inverse Problem}

A vital problem in the analysis of rocking rigid bodies is the reconstruction of a valid shape function
from a desired effective potential. Solving this inverse problem is not only a major step towards
the calculation of tautochrone shapes (see next section), but also a general tool to create rocking
rigid bodies with tailor-made dynamical properties.

To begin with, I restate the definition of the effective potential
\begin{eqnarray}
U(s) &=& g (v(s) - h)
\end{eqnarray}
where the potential is normalized such that $U(0) = 0$ holds for convenience. Instead of a result, $U(s)$ is now
considered as a free function and the shape function $z$ (or equivalently $w$ or $w_b$) is solved for.
Applying the inverse potential function $U^{-1}$ and deriving both sides of the equation with respect to $\theta$ yields
\begin{eqnarray}
g u_b &=& - F\big(g(v-h)\big) \sqrt{ |w_b|^2 +\Theta } \,. \label{invBase}
\end{eqnarray}
$F$ is an auxiliary function which maps the effective potential to its corresponding gradient:
\begin{eqnarray}
F(U(s)) &:=& \frac{dU(s)}{ds} \,.
\end{eqnarray}
In the most general case, $F$ is a multi-valued function which requires a separate inversion for each branch
and a careful assembly of these solutions into one or multiple global shapes. Therefore I assume a symmetric
effective potential with only on minimum at $s=0$ in the following to reduce the calculation down to a
single (e.g.\ $s\geq 0$) branch. Eq.~\ref{invBase} is an implicit equation for $w_b$ which can be directly solved for $u_b$
\begin{eqnarray}
u_b &=& - \frac{F\big(g(v-h)\big)}{\sqrt{g^2 - F\big(g(v-h)\big)^2}} \sqrt{v^2+\Theta} \label{ub_generic}
\end{eqnarray}
thus providing the solution parametrized as $w_b(v)$. Alternatively, depending on $F$, some other parametrization
$w_b(\lambda)$ might be more suitable. It remains to calculate the shape function $z$.
Rearranging Eq.~\ref{Eqdz_wb} suggests the Ansatz:
\begin{eqnarray}
z(\lambda) &=& i h - w_b(\lambda) \exp\left( i \theta(\lambda) \right) \,. \label{zAnsatz}
\end{eqnarray}
Re-inserting this Ansatz into Eq.~\ref{Eqdz_wb} provides the required differential equations
for $\theta(\lambda)$ and $b(\lambda)$:
\begin{eqnarray}
\frac{d\theta}{d\lambda} &=& - \frac{1}{u_b}   \frac{dv}{d\lambda} \label{theta_l_eq}\\
\frac{d b}{d\lambda}     &=& - \frac{1}{2 u_b} \frac{d|w_b|^2}{d \lambda} \,.
\end{eqnarray}
Both equations are manifestly invariant under change of parametrization.

\section{Tautochrone Shape} \label{SecTautochrone}

To solve the tautochrone shape problem, it is left to apply the inversion procedure from the previous section to an
isochronous effective potential $U(s)$. It is tempting to assume that only harmonic potentials are isochrone
potentials. But, in fact, the harmonic potential is only one solution among many other isochronous potentials.
Only if the solution space is restricted to symmetric $\mathrm{C}^2$ potentials it stands out as the sole solution \cite{bolotin2003Isochronous}.
Therefore I restrict the sought isochronous shapes to symmetric solutions only and continue with
the harmonic potential:
\begin{figure}
\centering
\includegraphics*[scale=.5, angle=270]{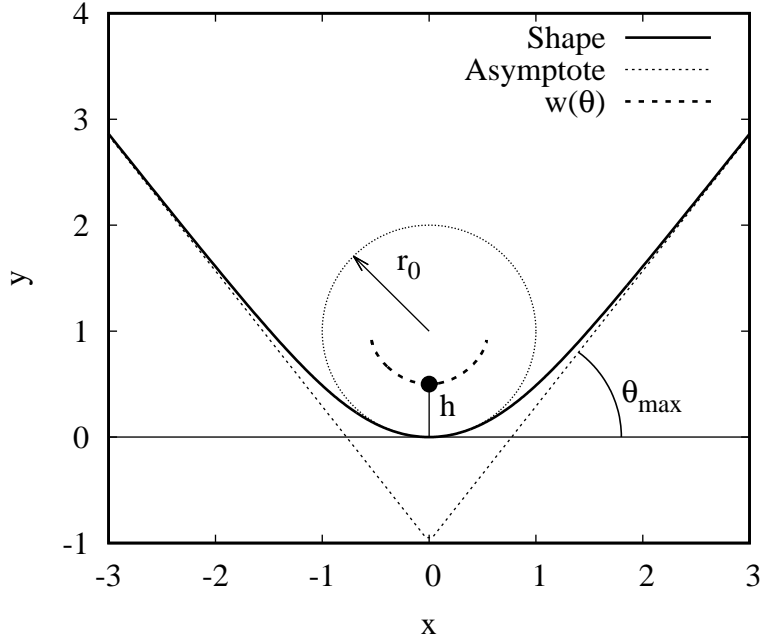}
\caption{\label{Demo}Tautochrone shape solution for $h/r_0 = 0.5$, $\eta = 0.6$ and $r_0=1$. The radius of curvature at the
origin and the asymptotes are highlighted by dashed lines. The thick dashed line marks the path $w(\theta)$ of the
center of mass in space-fixed coordinates.}
\end{figure}
\begin{eqnarray}
U(s) &=& \frac{1}{2} \omega_0^2 s^2 \\
F(U) &=& \omega_0 \sqrt{2 U} \,. \label{TautoF}
\end{eqnarray}
$U(s)$ needs to be consistent with the small angle approximation. Hence $\omega_0$ is identical to the frequency defined by Eq.~\ref{eq_w0}.
Inserting Eq.~\ref{TautoF} into Eq.~\ref{ub_generic} gives
\begin{eqnarray}
u_b^2 &=& \frac{v-h}{L +h - v} \left(v^2+\Theta\right) \,. \label{Eqtautochron}
\end{eqnarray}
By construction, the solution of the auxiliary parameter $s(t)$ is
\begin{eqnarray}
s(t) = s_0 \cos(\omega_0 t) \,.
\end{eqnarray}
The right hand side of Eq.~\ref{Eqtautochron} needs to be positive, thus the valid range of $v$ is restricted to the interval $[h,h+L]$.
Likewise, $|s(t)|$ is bounded by $s_{\max} = 2L$.
This motivates to use $L$ as the scale length of the problem at hand and to switch to a dimensionless set of parameters \cite{curtis1982dimensional}:
\begin{eqnarray}
\nu          &:=& \frac{v-h}{L} \\
\eta         &:=& \frac{h^2}{h^2+\Theta} \label{def_eta} \\
\delta       &:=& \frac{r_0-h}{h} \,.
\end{eqnarray}
The tautochrone shape solution is monostatic by construction. Thus it is possible to parametrize the shape function by $\nu$, though the
domain of $\theta$ needs to be restricted to $\theta \geq 0$ (there is a symmetric branch for $\theta \leq 0$).
All previously introduced parameters are readily expressed as dimensionless quantities:
\begin{eqnarray}
\tilde h      &=& 2\eta\delta \\
\tilde r_0    &=& 2\eta\delta (1+ \delta) \\
\tilde \Theta &=& 4\eta\delta^2(1-\eta) \,.
\end{eqnarray}
The tilde is a short-hand notation for the normalization by an appropriated power of $L$ (e.g.\ $\tilde h=h/L$).
Rewriting Eq.~\ref{Eqtautochron} using normalized quantities reads
\begin{eqnarray}
{{\tilde u}_b}^2 &=& \frac{\nu}{1-\nu} (\nu^2+4 \eta \delta( \nu + \delta)) \,.
\end{eqnarray}
Now all tools are in place for a detailed discussion of the properties of the tautochrone shape solution.
It is important to note that some tautochrone shapes actually ``roll'' along the underside of the support
plane, which is true for $\delta \in (-1,0)$ (i.e.\ $r_0<0$). For completeness, this special case is
discussed in Appendix~\ref{Appendix_rNegative}.
\begin{figure}
\centering
\includegraphics*[scale=.5, angle=270]{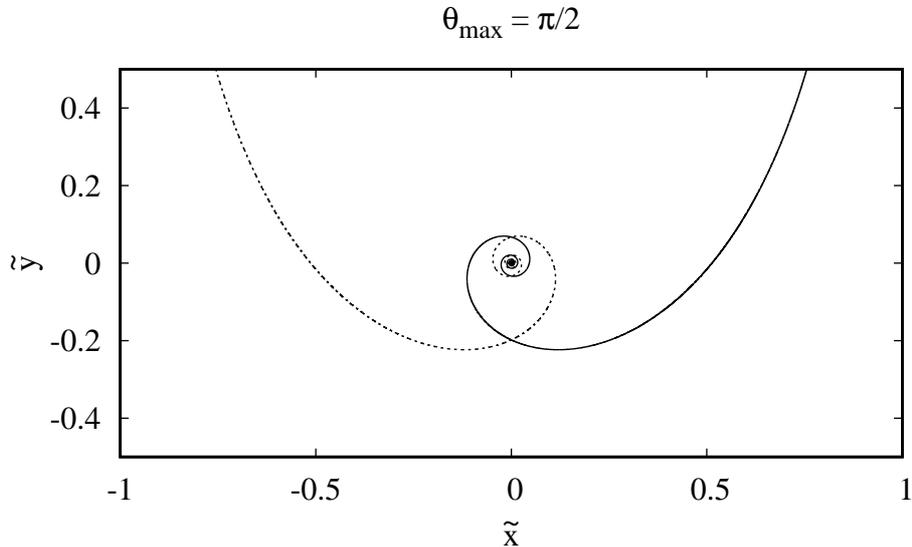}
\caption{\label{Spiral}Shape function limit for $\eta\delta^2 \rightarrow 0$. Plotted are $\theta>0$ (solid line) and $\theta <0$ (dashed line). }
\end{figure}
\subsection{Shape Function}

The rocking angle $\theta(\nu)$ is obtained by solving the differential equation~\ref{theta_l_eq} with initial condition $\theta(0)=0$:
\begin{eqnarray}
\theta(\nu)   &=& \int_0^{\nu} \sqrt{ \frac{1-\nu'}{\nu'({\nu'}^2+  4\eta \delta( \nu' + \delta))} } d\nu' \,.
\end{eqnarray}
In general, this integral can only be solved by means of elliptic integrals and is not representable by elementary functions.
Nonetheless, it is possible to obtain valuable insight into the characteristics of the resulting shape function by examining
the local radius of curvature (recall Eq.~\ref{r_dwb_db})
\begin{eqnarray}
r (\nu)       &=& \frac{r_0+L/2}{(1-\nu)^2} -L/2 \,. \label{rTautochrone}
\end{eqnarray}
$r (\nu)$ is a simple monotonously increasing function assuming infinity at $\nu=1$. Thus $z(\nu)$ asymptotically approaches
a straight line for $\nu \rightarrow 1$. The asymptote is readily obtained by formally expanding $\tilde z(\nu)$ at $\nu=1$:
\begin{eqnarray}
\tilde z(\nu) &=&  i \tilde h - i(1+ \tilde h)\exp( i \theta_{\max}) +\sqrt{\frac{1+ 2\tilde r_0}{1-\nu}} \exp( i \theta_{\max})
+\mathcal{O}(\sqrt{1-\nu}) \\
\theta_{\max} &=& \left. \theta \right|_{\nu=1}\,.
\end{eqnarray}
An alternative representation of the asymptote is
\begin{eqnarray}
y &=& \left(h - \frac{L+h}{\cos(\theta_{\max})}\right) + \tan(\theta_{\max})\, x \,.
\end{eqnarray}
Fig.~\ref{Demo} illustrates this transition from a circular arc of radius $r_0$ into a straight line running to infinity, which
is a characteristic of all tautochrone shapes. $\theta_{\max}$ is directly related to the winding number $\chi = \theta_{\max}/\pi +1/2$
defined with respect to the center of mass. Large $\theta_{\max}$ values result in self-intersecting shape functions. Though these
self-intersections do not pose any mathematical problem, they are a key challenge when actually building real solids representing these
solutions.
It is possible to study some properties of $\theta_{\max}(\eta,\delta)$ by considering special cases in the $(\eta,\delta)$
parameter space. At $\eta=1$, $\theta_{\max}$ can be expressed by elementary functions:
\begin{eqnarray}
\left. \theta_{\max} \right|_{\eta=1} &=& \pi \left| \sqrt{\frac{1}{2\delta}+1} -1\right|\,.
\end{eqnarray}
\begin{figure}
\centering
\includegraphics*[scale=.8, angle=0]{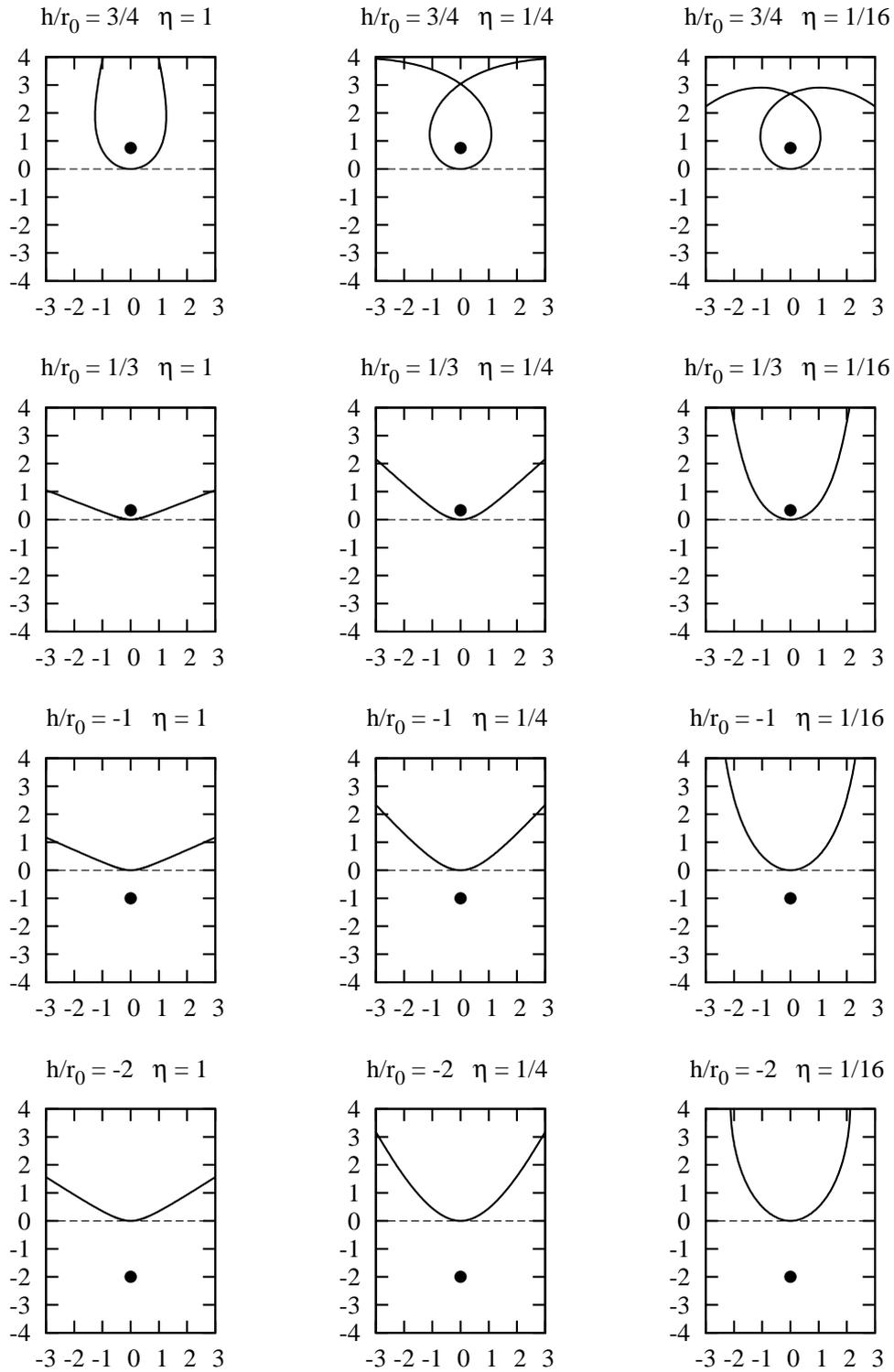}
\caption{\label{z_plot}Tautochrone shape solutions for selected values of $h/r_0$ and $\eta$. $r_0=1$ for all plots. The center of mass is marked by a black dot.}
\end{figure}

As $h$ approaches $r_0$, the shape function winds ever tighter around the center of mass. On the contrary, this is not
true for $h \rightarrow -\infty$ (i.e.\ the rocking pendulum limit). The rocking angle is bound by $ \pi(1-\sqrt{1/2}) \approx 52.7^{\circ}$
in this case.
Small values of $\eta \ll \min(1, 1/\delta^2)$ permit the approximation
\begin{eqnarray}
\theta_{\max} & = & \int_0^{\infty} \frac{1}{\sqrt{\nu(\nu^2+4\eta\delta^2)} } d\nu +\mathcal{O}(\delta\sqrt{\eta}) + \mathcal{O}(\sqrt{\eta}) \\
& \approx & \frac{2K(1/\sqrt{2})}{\sqrt[4]{4\eta\delta^2} } \,.
\end{eqnarray}
$K(\cdot)$ is the complete elliptic integral of the first kind \cite{abramowitz1972handbook}.
$\eta\delta^2$ is an overall good indicator of the winding number of a shape solution. A compilation of solutions in
Fig.~\ref{z_plot} illustrates the trend towards larger winding numbers as this parameter increases. The two center rows
corresponding to $\delta =\pm 2$ further underline this finding. Though the center of mass is located once above and once
below the supporting plane, both sets of tautochrone shapes are remarkably similar.

In particular intriguing is the limit $\eta\delta^2 \rightarrow 0$. At first, it seems that
a well defined limit does not exist, as the rocking angle integral diverges. However, it is possible
to introduce $\theta_{\max}$ as a free parameter to obtain a well defined family of solutions:
\begin{eqnarray}
\theta(\nu)    &=& - 2\sqrt{ \frac{1-\nu}{\nu}} + \arccos(2\nu-1) + \theta_{\max} \\
\tilde z(\nu) &=&  \left( \frac{\nu^{3/2}}{\sqrt{1-\nu}}   - i\nu \right) \exp\left(i\theta(\nu)\right)\,.
\end{eqnarray}
Since $\tilde h=0$ holds in this limit as well, $\theta_{\max}$ drops out of Eq.~\ref{w_phiDefEq}
and the center of mass moves independently from $\theta_{\max}$ on the same path.
Fig.~\ref{Spiral} illustrates the solution for $\theta_{\max} = \pi/2$. Close to the origin the rocking
angle varies as $\theta \approx -2|\tilde z|^{-1/2}$ which is a general Archimedean spiral with exponent $-1/2$.
The next section proves another fascinating property: The center of mass moves along a cylcoid when unrolling
this curve.
 
\subsection{Center of Mass}

The tautochrone center of mass path is the solution of the differential equation
\begin{eqnarray}
\frac{d \tilde u }{d \nu}   &=& (2 \eta \delta + \nu)\sqrt{ \frac{1-\nu}{\nu(\nu^2+  4\eta \delta( \nu + \delta))} } \,. \label{Equv_mass}
\end{eqnarray}
As for the rocking angle, the most general solution needs to be expressed in terms of elliptic integrals. An inspection of the right hand side
provides the bound $|\tilde u| \leq \pi/2$ for all admissible values of $(\eta,\delta)$. Further insight might be obtained by considering
the case of a vanishing moment of inertia $\tilde \Theta = 0$, which implies either $\eta=0$, $\eta=1$ or $\delta=0$. With all the
mass concentrated in a single point, the rocking rigid body is equivalent to a point mass moving along a fixed path and the
solution should be identical to the classical tautochrone. Indeed, formally solving Eq.~\ref{Equv_mass} at $\tilde \Theta = 0$ recovers the well known inverted cycloid:
\begin{eqnarray}
\tilde u   &=& \frac{1}{2}( \sin(\varphi) +\varphi) \\
\tilde v   &=& \tilde h + \frac{1}{2} (1-\cos(\varphi)) \\
\tilde  u_b &=& - \tan(\varphi/2) |\tilde v| \\
\varphi & \in & [-\pi,\pi] \,.
\end{eqnarray}
\newpage
\begin{figure}[!ht]
\centering
\includegraphics*[scale=.8, angle=0]{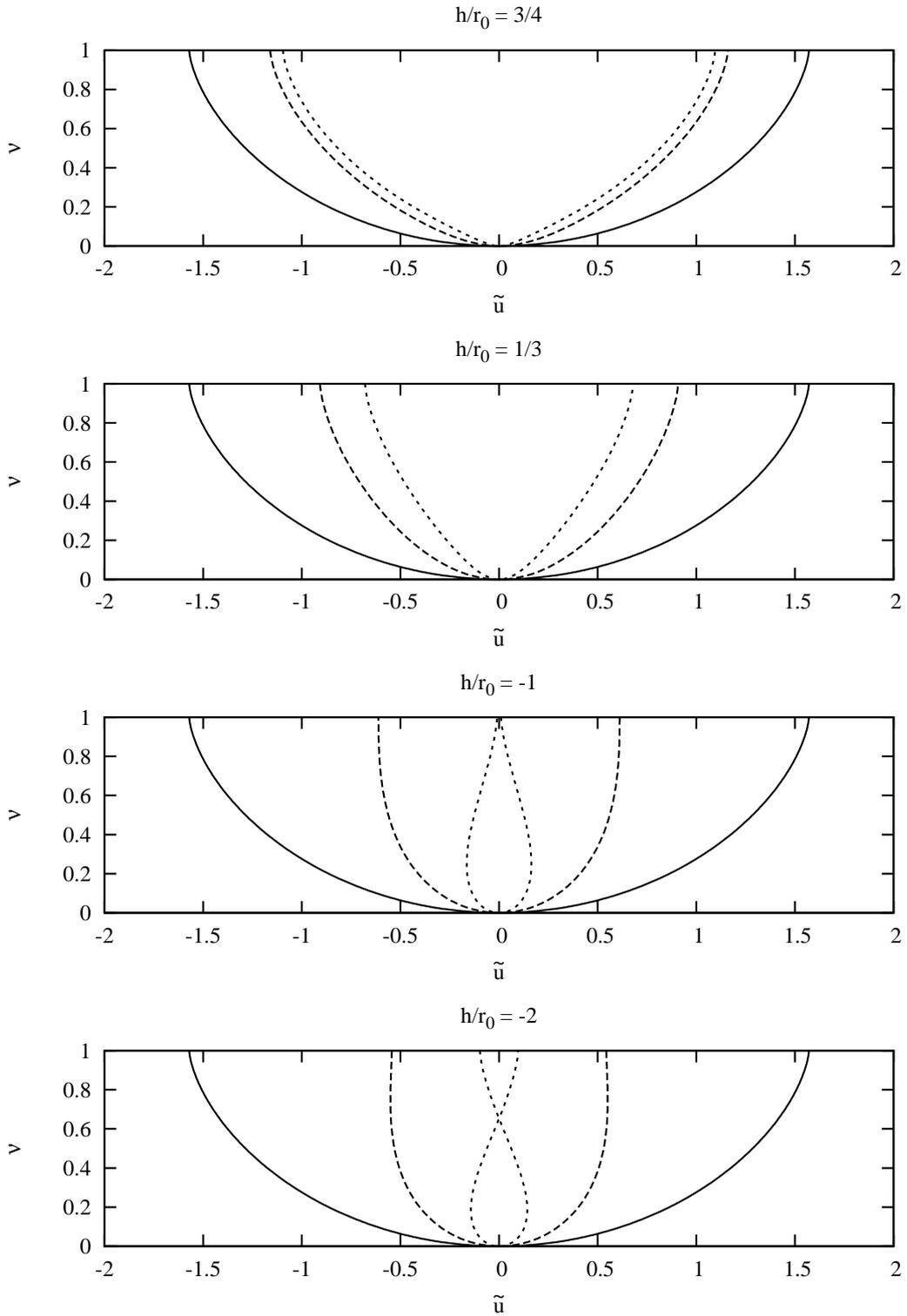}
\caption{\label{w_plot}Center of mass paths for selected values of $h/r_0$ and $\eta$.
Plotted are $\eta=1$ (thick line), $\eta=1/4$ (dashed line), $\eta=1/16$ (dotted line). $r_0=1$ for all plots.}
\end{figure}
\newpage
The cycloid parameter $\varphi$ has a simple interpretation for maximal rocking amplitude $\tilde \nu_A=1$.
By considering the corresponding solution $s_{\max}(t)$ of the auxiliary parameter
\begin{eqnarray}
s_{\max}(t) &=& 2L \sin(\varphi(t)/2)
\end{eqnarray}
the cycloid parameter is identified as being proportional to the elapsed time: $\varphi = 2\omega_0 t$. Though other parameter
values $(\delta, \eta)$ do not allow for an elementary analytical solution, most center of mass paths still resemble a
squeezed version of the inverted cycloid, as illustrated by selected center of mass paths in Fig.~\ref{w_plot}.
Whenever the center of mass crosses the support plane, $v$ changes its sign and the center of mass pass path bends inwards,
eventually leading to a self-intersection (see Fig.~\ref{w_plot}, lower panels).

\section{Slipping}

The analysis assumed so far a perfect rolling motion without any slipping. Though this is appropriate for a
theoretical study, real bodies might in fact slip or even lift of the support plane \cite{gomez2012jumping}. A common physical model
is to employ the static friction coefficient to formulate a physical no-slip condition based on the
constraint acceleration $a$ acting at the contact point:
\begin{eqnarray}
|a_{\parallel}| < \mu a_{\perp} \,.  \label{NoSlipCond}
\end{eqnarray}
$a_{\parallel}$ is the component parallel to the support plane, $a_{\perp}$ is the perpendicular component and $\mu$
is the static friction coefficient. In fact, Eq.~\ref{NoSlipCond} precludes jumps as well since an upward acceleration implies
negative values of $a_{\perp}$.
Inserting the tautochrone solution into the expressions obtained in Section~\ref{SecConstForce} gives
\begin{eqnarray}
a_{\perp}    &=&  g (1-\nu) + \frac{1}{2L} \big( |w_b|^2 + \Theta \big) {\dot \theta}^2 \\
a_{\parallel} &=&  \frac{u_b v }{v^2+\Theta}a_{\perp}  - \frac{u_b\Theta}{v^2+\Theta} {\dot \theta}^2 \,.
\end{eqnarray}
By  invoking the conversation of energy again, $\dot \theta$ can be expressed as
\begin{eqnarray}
{\dot \theta}^2 &=& 2gL \frac{\nu_A-\nu}{|w_b|^2 +\Theta}
\end{eqnarray}
which simplifies the constraint acceleration to
\begin{eqnarray}
a_{\perp}     &=&  g (1+\nu_A - 2\nu)\\
a_{\parallel} &=&  \frac{u_b v }{v^2+\Theta}\left(a_{\perp}  - 2g\frac{\Theta}{|w_b|^2 +\Theta} \frac{\nu_A-\nu}{\nu +\tilde h}  \right) \,.
\end{eqnarray}
$a_{\perp}$  is always positive, therefore tautochrone shapes are not subjected to jumping for all admissible values of $\nu_A$.
The constraint acceleration is both a function of $\nu_A$ and $\nu$. Thus the no-slip condition needs to be
validated for the entire rocking motion $\nu \in [0,\nu_A]$.
This defines the set $S_R$ of rocking amplitudes which preclude slipping:
\begin{eqnarray}
S_R := \{\nu_A \,|\, \forall \nu \in [0,\nu_A] : |a_{\parallel}| < \mu a_{\perp} \} \,. \label{SRset}
\end{eqnarray}
The upper bound $\nu_{\mathrm{slip}} := \sup(S_R)$ is of particular interest, as it defines the range of the shape function that
is accessible by a physical rocking motion. In general, Eq.~\ref{SRset} does not admit a simple closed-form solution for
$\nu_{\mathrm{slip}}$. However, imposing $\tilde \Theta=0$ simplifies Eq.~\ref{SRset} to
\begin{eqnarray}
|u_b| < \mu |v|.
\end{eqnarray}
\begin{figure}
\centering {
\includegraphics*[scale=.8, angle=0]{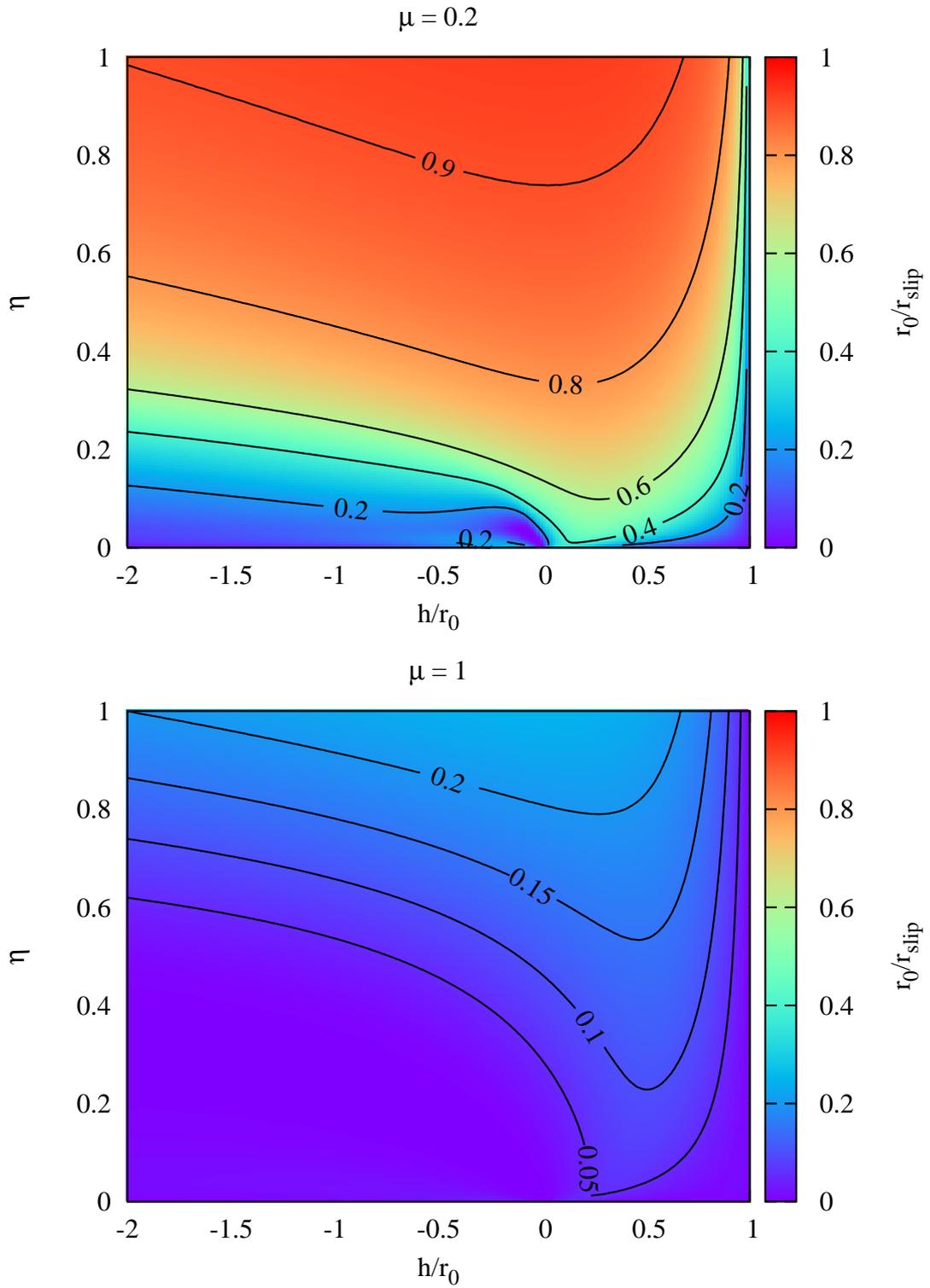}}
\caption{\label{r_slip}$r_0/r_{\mathrm{slip}}$ at the onset of slipping for two selected values of $\mu$.}
\end{figure}

This expression can be further reduced to a concise analytical solution
\begin{eqnarray}
\left. \nu_{\mathrm{slip}} \right|_{\tilde \Theta=0} &=& \frac{\mu^2}{1+\mu^2} \,.
\end{eqnarray}
Though this solution is only strictly valid for $\tilde \Theta=0$, the general trend applies to other values of
$\tilde \Theta$ as well: Larger values of the friction coefficient $\mu$ allow the rigid body
to explore a larger range of its shape function. However, $\nu_{\mathrm{slip}}$ does not provide
an immediate indication as to what range of the shape function is accessible. All tautochrone shapes are well
approximated by a circular arc for sufficiently small values of $\nu_{\mathrm{slip}}$. The practical relevance of the
derived tautochrone shapes is therefore characterized best by considering the ratio $r_0/r(\nu_{\mathrm{slip}})$,
which quantifies the maximal deviation from a simple circular shape. Making use of Eq.~\ref{rTautochrone}
provides the equation
\begin{eqnarray}
r_0/r_{\mathrm{slip}} &=& \frac{4\eta\delta(1+\delta) (1-\nu_{\mathrm{slip}})^2 } { 4\eta\delta(1+\delta) + 1 - (1-\nu_{\mathrm{slip}})^2 } \,. \label{Eqrslip}
\end{eqnarray}
Neglecting the dependence on $\nu_{\mathrm{slip}}$, this equation favors small values of $\eta$ and $\delta$.
This trend can be further substantiated by numerically solving Eq.~\ref{SRset} for selected values of $\mu$.
Fig.~\ref{r_slip} presents two choices, one being $\mu = 0.2$, a value not uncommon for most everyday materials like
wood or plastic \cite{onorato2014librational}, and a second choice $\mu = 1$ which requires more sticky surfaces like rubber.
Results clearly favor large $\mu$ values, though viable regions in the parameter space still exist for $\mu = 0.2$.

\section{Brachistochrone Shape} \label{SecBrachistochrone}

A brachistochrone shape minimizes the time $T$ required to roll from a given starting position back into its equilibrium
position. Using the previously introduced notation $T$ is defined as
\begin{eqnarray}
T &=& \int_{0}^{\theta_{\max}} - \frac{d\theta}{\dot \theta}  \\
&=& \int_{0}^{\theta_{\max}} \sqrt {\frac{|w_b|^2 +\Theta}{2g(v_{\max}-v)}} d\theta
\end{eqnarray}
where $\theta_{\max}$ sets the initial rocking angle and the integration is performed over the $\theta \geq 0$ branch.
In agreement with the tautochrone problem, both the moment of inertia $\Theta$ and the height of the center of mass $h$
are considered as fixed parameters.
It is beneficial to change to $v$ as integration variable by employing Eq.~\ref{du_dv}:
\begin{eqnarray}
T &=& \int_{v_{\min}}^{v_{\max}} \sqrt {\frac{1 + \left(\frac{du}{dv}\right)^2(1 +\Theta/v^2)}{2g(v_{\max}-v)}} dv \,. \label{variat_uv}
\end{eqnarray}
Eq.~\ref{variat_uv} also provides a proper definition the ambiguous term ``starting position'' as the initial position of the center of mass
in space-fixed coordinates. The integrand does not depend explicitly on $u$, thus reducing the Euler-Lagrange equation to
\begin{eqnarray}
\frac{du}{dv} \frac{1 +\Theta/v^2}{\sqrt {(1 + \left(\frac{du}{dv}\right)^2(1 +\Theta/v^2))(v_{\max}-v)}}   &=& \kappa
\end{eqnarray}
where $\kappa$ is a free integration constant carrying the same sign as $du/dv$.
This equation can be solved for $du/dv$:
\begin{eqnarray}
\left(\frac{du}{dv}\right)^2 &=& \frac{v_{\max}-v}{\left((1+\Theta/v^2)/\kappa^2 - (v_{\max}-v)\right)(1+\Theta/v^2)} \,. \label{Brachis_uv}
\end{eqnarray}
The corresponding solution $u_b$ is
\begin{table}
\end{table}
\begin{figure}[!ht]
\centering { \includegraphics*[scale=.4, angle=270]{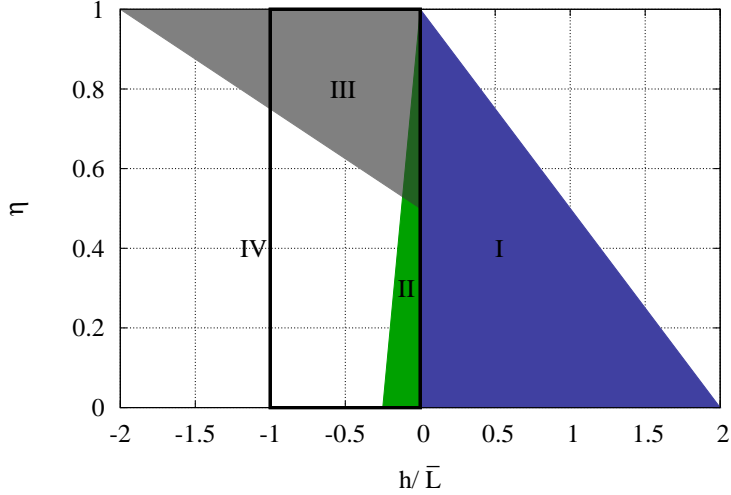} }
\caption{\label{brachisScetch} Classification of the brachistochrone shape solutions. See
Section~\ref{SecBrachistochrone}
for a detailed description of the different regions. }
\end{figure}
\begin{eqnarray}
u_{b}^2 &=&  \frac{ (1+\Theta/v^2) / \kappa^2 -(v_{\max}-v)}{v_{\max}-v}(v^2+\Theta) \,. \label{ub_brachis}
\end{eqnarray}
The constant $\kappa$ is determined by the boundary values of $u(v)$. By using the definition of the center of mass height $u_b(h)=0$,
it is possible to obtain a more meaningful expression for $\kappa$:
\begin{eqnarray}
\kappa^2 &=& \frac{1}{\eta \bar L} \,.
\end{eqnarray}
$\eta$ is defined as in Eq.~\ref{def_eta} and $\bar L$, in analogy to the tautochrone solution, is defined as
\begin{eqnarray}
\bar L &:=& v_{\max}-h \,.
\end{eqnarray}
Employing both $\kappa$ and $\bar L$ simplifies Eq.~\ref{ub_brachis} to
\begin{eqnarray}
u_b^2 &=&  \frac{v-h}{\bar L+h-v}(v^2+\Theta) \left( 1 - \frac{\bar L(1-\eta)}{v^2}(v+h) \right) \label{Brachis_ub}
\end{eqnarray}
which is almost identical to the corresponding equation of the tautochrone shape (compare Eq.~\ref{Eqtautochron}).
However, the additional factor in Eq.~\ref{Brachis_ub} is only negligible in the limit $ |h|\gg \bar L$, where
the brachistochrone shape converges to the tautochrone shape solution. Finite values of $h$ require a much more
careful analysis, and, as we will see, the resulting brachistochrone shapes are in remarkable contrast to the
simplicity of the tautochrone solution. The first complication arises from the fact that the right hand side of
Eq.~\ref{Brachis_ub} might become negative, and hence both $du/dv$ and $u_b$ are undefined. Though this precludes
the existence of a global minimizer of the variational problem, it is possible to join valid branches of the
solution to obtain a minimizer on a reduced function space. Returning to the original variational Eq.~\ref{variat_uv}
motivates to replace undefined values by $du/dv = 0$ for which the integrand assumes its minimum value. While this
implies $u_b^2 = \infty$, this apparently strange result has a simple physical interpretation: The rigid body
actually does not rotate at all but performs a free fall at velocity $\sqrt{2g(v_{\max}-v)}$.

Another remarkable property stems from the fact that $u_b$ is singular at $v=0$, implying that the
brachistochrone shape decomposes into two separate parts. The rocking rigid body switches from one branch
of the shape solution to another as the center of mass crosses the support plane.

Finally, calculating the local radius of curvature $r_0$ at the presumed equilibrium contact point $v=h$ gives
\begin{eqnarray}
r_0    &=& h + \frac{h}{2\eta}\left( \frac{h}{\bar L} - 2(1-\eta)\right) \,. \label{r0_brachis}
\end{eqnarray}
An inspection of $r_0$ reveals the somewhat disturbing property that the brachistochrone shape solution
might violate the stability criterion $r_0 > h$. All these peculiarities can be consolidated by a detailed
analysis of Eq.~\ref{Brachis_ub}, which is presented in Appendix~\ref{AppendixZeroth} for clarity.
The result of this analysis is summarized best in the ($h/\bar L$, $\eta$) parameter plane (Fig.~\ref{brachisScetch}).
The highlighted regions are:
\begin{enumerate}
\item[I)]
The brachistochrone shape initially rolls along the support plane, followed by a free-fall down to $v=h$. Though
Eq.~\ref{r0_brachis} technically violates the stability criterion $r_0 > h$ in this case, this is only a  mathematical
artefact as the shape solution and therefore the local radius of curvature at $v=h$ is not defined.
      
\item[II)] The brachistochrone shape initially rolls along the support plane, performs an intermittent free-fall and
continues the rolling motion on a second branch of the shape solution.
      
\item[III)] The local radius of curvature at $h=v$ is negative, thus indicating a rigid body rolling along
the underside of the support plane.

\item[IV)]
The center of mass crosses the support plane and switches over to a second branch of the shape solution.
\end{enumerate}
Some of these regions partly overlap, which generates a rich variety of brachistochrone shape solutions.
Sadly, this diversity largely represents extremely difficult-to-construct shape solutions, and only
tautochrone-like shapes appear to be viable solutions.

\section{Discussion}

Rocking rigid bodies constitute an intriguing classical mechanics problem. This work
presented a self-contained analysis of their dynamical properties, which can be applied to
a broad range of rocking rigid bodies. In particular, this work focused on the
derivation of two novel shape solutions: the tautochrone and the brachistochrone shape.
Both shapes form a two-parametric family of solutions, parametrized by the center of mass
height $h$ and the moment of inertia $\Theta$. By construction, tautochrone shapes
exhibit a much more regular behavior than their brachistochrone counterparts.
Moreover, tautochrone shapes offer a new alternative solution to Huygen's cycloid approach
to build an isochronous pendulum.

Although both the tautochrone and the brachistochrone shape are sufficiently described by this
analysis, there remains an interesting challenge: Is it possible to demonstrate their properties
by actually building physical representations of these shapes? A first analysis
of the no-slip condition does not prohibit their construction, but the available parameter
space is narrowed down to shapes with a comparably large moment of inertia (i.e.\ small $\eta$ values).
In addition, both $h$ and $\Theta$ are coupled through the mass distribution of the rigid body,
which calls for an optimal design in terms of simplicity and elegance. This problem is left for future work.

\begin{appendix}

\newpage
\section{Illustrative Examples} \label{AppendixConical}

An elegant set of rigid body shapes is obtained if the center of mass in comoving coordinates $w_b$
is required to move along a conic section with eccentricity $\varepsilon_b$. This Ansatz allows for
a complete set of analytic solutions for $\theta$, $r$, $w$ and consequently $z$. Furthermore, the
center of mass moves along a conic section with eccentricity $\varepsilon$ in space-fixed coordinates as well.
Given $h$ and $r_0$, two different solutions (either hyperbolas or ellipses) are obtained, parametrized by a
suitably chosen parameter $\lambda$:
\begin{center}
\begin{tabular}{c|r|r}
& $h>0, r_0>h$  & $h<0, r_0>h$ \\ \hline
$\theta$         & $\lambda/ \sqrt{r_0/h-1}$                  & $\lambda/ \sqrt{1-r_0/h}$  \\ \hline
$u_b$            & $  -h\sqrt{r_0/h-1}\sinh(\lambda)$         & $h\sqrt{1-r_0/h}\sin(\lambda) $  \\ \hline
$\varepsilon_b^2$ & $\frac{r_0}{h}$                            & $\frac{r_0}{r_0-h}$ for $r_0 \geq 0$ \\
&                                            & $\frac{r_0}{h}$ for $r_0 < 0$ \\ \hline
$u$             & $\frac{h}{\sqrt{r_0/h-1}}\sinh(\lambda) $  & $\frac{h}{\sqrt{1-r_0/h}} \sin(\lambda) $  \\
$v$             & $h \cosh(\lambda) $                        & $h \cos(\lambda) $  \\ \hline
$\varepsilon^2$   & $\frac{r_0}{r_0-h}$                        & $\varepsilon_b^2$ \\ \hline
$r$              & $r_0\cosh(\lambda) $                       & $r_0\cos(\lambda) $  \\
\end{tabular}
\end{center}
Shape solutions $z(\lambda)$ for $h>0$ are a sum of two counter-rotating logarithmic spirals, while shape solutions
for $h<0$ are epicycloides $(r_0<0)$ and hypocycloids $(r_0>0)$.

\begin{figure}[h]
\centering {
\includegraphics*[scale=.54, angle=270]{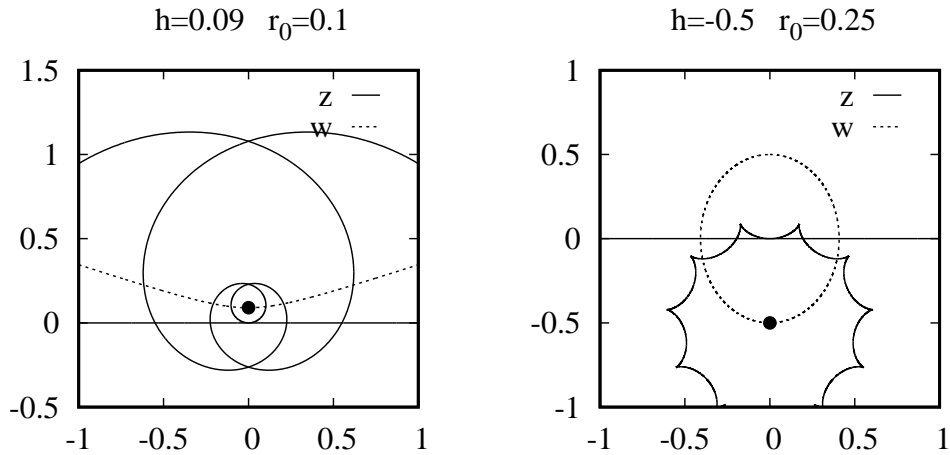}}
\caption{\label{examples} Sample curves showing the shape solution (solid line) and center of mass path (dashed line) for two
selected parameter sets $(h,r_0)$. }
\end{figure}

\newpage
\section{Tautochrone Shape} \label{Appendix_rNegative}

\begin{figure}[h]
\centering {
\includegraphics*[scale=.4, angle=270]{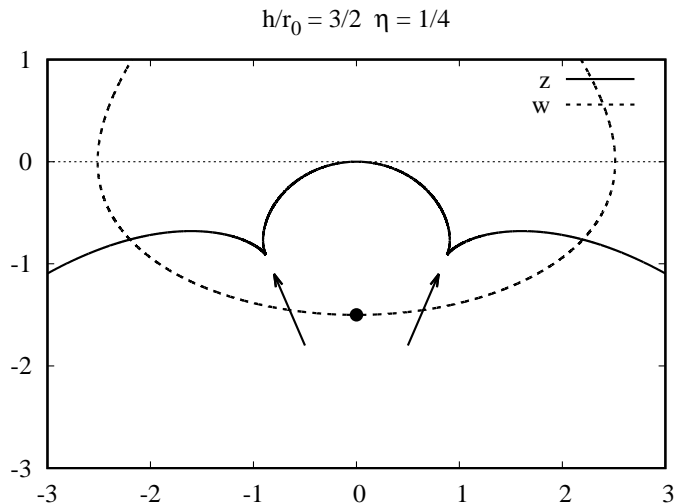}}
\caption{\label{rnegScetch} Tautochrone shape solution for $r_0=-1$. The black dot is the center of mass and the dashed line
is the center of mass path. Arrows indicate the cusp-like features where $r=0$. }
\end{figure}

All properties discussed in Section~\ref{SecTautochrone} apply equally well to tautochrone shape solutions
with negative values of $r_0$. In addition, the corresponding shape functions exhibit cusp-like features similar to
epicycloids. This can be seen most easily  by considering the tangent vector of the shape solution
\begin{eqnarray}
\frac{dz}{d\theta} &=& r \exp(i\theta) \,.
\end{eqnarray}
The rocking angle $\theta$ is continuous and differentiable at $r=0$. Hence the tangent vector abruptly inverts
its direction at $r=0$, leading to the cusps seen in Fig.~\ref{examples} (right panel) and Fig.~\ref{rnegScetch}. The local radius of curvature vanishes at
\begin{eqnarray}
\tilde v_{(r=0)} &  = &  1 + 2\eta\delta -\sqrt{4  \eta\delta(1+\delta)+1} \,.
\end{eqnarray}
A second interesting feature emerges when the center of mass may cross the support plane, which requires $\tilde h \in (-1,0)$.
By approximating $\tilde u_b$ at $\tilde v=0$
\begin{eqnarray}
{\tilde u}_b^2 & \approx & C \left({\tilde v}^2 + \tilde\Theta\right) \\
C &:=& \frac{-\tilde h}{1 + \tilde h}
\end{eqnarray}
it is possible to derive an analytical solution for $w_b$
\begin{eqnarray}
{\tilde u}_b & \approx & -\sqrt{C\tilde \Theta} \cosh( \sqrt{C} (\theta - \theta_0) ) \\
\tilde v    & \approx &\,\quad \sqrt{\tilde\Theta} \sinh( \sqrt{C} (\theta - \theta_0) )
\end{eqnarray}
which is equivalent to the first example presented in Appendix~\ref{AppendixConical}.
In the limit $\eta \rightarrow 1$ both $\tilde \Theta$ and $\tilde v_{(r=0)}$ tend to zero and the
cusp turns into a joining point of two infinite, logarithmic spiral-like curves. However, the divergence is
rather slow as the number of spiral revolutions is $\mathcal{O}(\ln(1-\eta))$.

\section{Brachistochrone Shape} \label{AppendixZeroth}

$u_b^2$ as defined by Eq.~\ref{Brachis_ub} might assume negative values, depending on $h/\bar L$ and $\eta$.
To characterize the values ($h/\bar L,\eta$) which lead to negative values, it is sufficient to study the
roots of the polynomial
\begin{eqnarray}
P(v) &:=& v^2 - \bar L(1-\eta)(v+h) \,.
\end{eqnarray}
The roots of this quadratic polynomial are
\begin{eqnarray}
\frac{v_{1,2}}{\bar L } &=& \frac{1-\eta}{2} \pm \sqrt{ \frac{(1-\eta)^2}{4} + \frac{h}{\bar L}(1-\eta) } \,.
\end{eqnarray}
At $v_{\max}$, $P$ evaluates to
\begin{eqnarray}
P(v_{\max}) &=& P(h+\bar L) \\
&=& (h+\bar L\eta)^2 + \eta(1-\eta){\bar L}^2
\end{eqnarray}
which is a positive quantity for all admissible values ($h/\bar L, \eta$). Therefore $P$ assumes negative
values on the interval $[h,h+\bar L]$ if and only if $P$ has at least one root in this interval.
A necessary condition is that both roots are real, thus requiring
\begin{eqnarray}
\frac{h}{\bar L} & \geq & -  \frac{1-\eta}{4} \,.
\end{eqnarray}
A detailed analysis of both roots provides the intervals
\begin{eqnarray}
\begin{array}{lclclcl}
v_1 & \in & [h, h + \bar L] & \Leftrightarrow & h/\bar L &\in& [-(1-\eta)/4,\,0] \\
v_2 & \in & [h, h + \bar L] & \Leftrightarrow & h/\bar L &\in& [-(1-\eta)/4,\,2-2\eta]
\end{array} \,.
\end{eqnarray}
The intersection of both intervals, corresponding to a roll-drop-roll sequence, defines region II in Fig.~\ref{brachisScetch},
whereas region I corresponds to a single root $v_2$ and a roll-drop sequence.

\end{appendix}

\newpage

\nocite{*}
\bibliographystyle{plain}
\bibliography{RockingRigid}

\end{document}